\documentclass[11pt]{article}
\usepackage{amsmath}
\usepackage{amsfonts}
\usepackage{graphics}
\usepackage{graphicx}
\usepackage{ragged2e}
\usepackage{float}
\usepackage{latexsym}
\usepackage{amsthm}
\usepackage{cite}
\usepackage{amssymb}
\usecounter{enumi}
\newtheorem{Lemma}{Lemma}[section]
\newtheorem{Theorem}[Lemma]{Theorem}
\newtheorem{Example}[Lemma]{Example}
\newtheorem{Remark}[Lemma]{Remark}
\newtheorem{Corollary}[Lemma]{Corollary}
\newtheorem*{Proof}{Proof}

\numberwithin{equation}{section}

\def\be{\begin{eqnarray}} \def\ee{\end{eqnarray}} \def\k{\kappa}
  \def\({\left(} \def\){\right)}
\def\bc{\begin{center}} 
\def\ec{\end{center}}  

\def\bey{\begin{eqnarray*}}\def\eey{\end{eqnarray*}}
\begin{document}
\title{On a class of third order differential equations describing pseudospherical or spherical surfaces}

\author{ Mingyue Guo$^{1}$, Jing Kang$^{1,2,3}$, Zhenhua Shi$^{1,2}$, Zhiwei Wu$^{4}$
\vspace{4mm}\\
$^{1}$\small{School of Mathematics, Northwest University, Xi'an 710069,  P.R. China}\\
$^{2}$\small {Center for Nonlinear Studies, Northwest University, Xi'an 710069, P.R. China}\\
$^{3}$\small {Shaanxi Key Laboratory of Mathematical Theory and Computation of Fluid Mechanics,} \\
 \small {Northwest University, Xi'an 710069, P.R. China}\\
 $^{4}$\small {School of Mathematics (Zhuhai), Sun Yat-sen University, Zhuhai, 519082 , P.R. China}}

\date{}
\maketitle
\begin{minipage}{140mm}

\begin{abstract}
In this paper, we study third order nonlinear partial differential equations which describe surfaces of constant curvature. From the flatness of connection 1-forms, we present a classification of equations with the type $u_t - u_{xxt} = \lambda u^2 u_{xxx} + G(u, u_x, u_{xx}) (\lambda\in\mathbb{R})$, which describe pseudospherical or spherical surfaces. We show that series of typical soliton equations belong to certain subclass, such as the generalized Camassa-Holm equation, which gives a geometric explanation to these equations.
\end{abstract}

\end{minipage}

\section{Introduction}

\indent \indent The concept of differential equations which describe pseudospherical surfaces was first introduced by Chern and Tenenblat \cite{chern1986}, inspired by Sasaki's observation \cite{sasaki1979}. They noted that all the soliton equations in $1+1$ dimension, which can be solved by the AKNS $2\times2$ inverse scattering method, describe pseudospherical surface. A classical example of this is the sine-Gordon (SG) equation, first discovered by Bour \cite{bour1891}.

The study of differential equations describing pseudospherical or spherical surfaces has both theoretical importance and practical relevance in the fields of mathematics and physics \cite{cavalcante1988}. These equations can be studied by using geometric methods, especially for their integrability conditions, which connect to $\mathfrak{sl}(2, \mathbb{R})$-valued or $\mathfrak{su}(2)$-valued linear problems, allowing their solutions to be obtained via the inverse scattering method \cite{ablowitz1974}.

Chern and Tenenblat not only developed a comprehensive approach for understanding these equations, but also built a classification method for differential equations of the form $ u_t = F(u, u_x, \ldots, \partial^k u / \partial x^k) $ \cite{chern1986}. This classification was extended to a variety of soliton equations, including the Liouville equations \cite{jorge1987}, sinh-Gordon equation \cite{rabelo1989}, KdV equation, modified KdV equation \cite{rabelo1992}, and etc..

In 1995, Kamran and Tenenblat refined this classification by eliminating the priori assumptions regarding the presence of a spectral parameter \cite{kamran1995}. This work was further generalized by Reyes in a succession of subsequent papers \cite{reyes1998,reyes2002,reyes2003,reyes2005,reyes2005b, reyes2006,reyes2011}. Additionally, Ding and Tenenblat \cite{ding2002} researched a differential system describing pseudospherical and spherical surfaces, including the nonlinear Schr\"odinger equation (NLS) and the Heisenberg Ferromagnet model.

This paper is concerned with the classification of the following third order partial differential equations of the form
\begin{equation}\label{1.1}
u_t - u_{xxt} = \lambda u^2 u_{xxx} + G(u, u_x, u_{xx}), \quad \lambda \in \mathbb{R},\quad G\neq0,
\end{equation}
which describe pseudospherical or spherical surfaces. The classification problem corresponds to determine (\ref{1.1}) admitting 1-forms
\begin{equation}\label{1.2}
\omega_i = f_{i1}\,dx + f_{i2}\,dt, \quad 1 \leq i \leq 3,
\end{equation}
where coefficient functions $\displaystyle f_{ij}=f_{ij}(u,u_x,\ldots,  u_x^{(k)})$ satisfy a particular system of equations.

Due to the cubic nonlinearity terms, the exhaustive analysis is relatively challenging. And it is imperative to impose further assumptions on coefficient functions $f_{ij}$. In our quest to ascertain what such an assumption should be, we drew substantial inspiration from the observation that the Camassa-Holm (CH) equation
 \begin{equation}\label{1.3}
u_t - u_{xxt} =u u_{xxx} + 2 u_x u_{xx} - 3 u u_x,
\end{equation}
describes pseudospherical surfaces with coefficient functions \cite{silva2015}
\begin{align*}
f_{11}&=u-u_{xx}+\frac{\eta^2}{2}-1,\quad &f_{12}&=-u\left(u-u_{xx}+\frac{\eta^2}{2}\right)\pm\eta u_x-\frac{\eta^2}{2}+1,\\
f_{21}&=\eta,\quad &f_{22}&=-\eta u\pm u_x-\eta,\\
f_{31}&=\pm\left(u-u_{xx}-\frac{\eta^2}{2}\right),\quad &f_{32}&=\pm u\left(u-u_{xx}+\frac{\eta^2}{2}\right)\pm\eta u_x\pm u\pm\frac{\eta^2}{2}.
\end{align*}
The crucial insight is that these coefficient functions satisfy the auxiliary condition
\begin{equation}\label{1.4}
f_{p1} = \mu_p f_{11} + \eta_p, \quad \mu_p, \eta_p \in \mathbb{R}, \quad p = 2, 3,
\end{equation}
which is consistently applied throughout our work. Such an auxiliary condition, in addition to extending the case $f_{21} = \eta$,  significantly simplifies the classification of differential equations that describe pseudospherical or spherical surfaces of the form (\ref{1.1}). This condition was first used by Gomes Neto \cite{neto2010} in the study of fifth order equations of the from $u_t = u_{xxxxx} + G(u, u_x, u_{xx}, u_{xxx}, u_{xxxx})$, and by Catalano Ferraioli and Tenenblat \cite{ferraioli2014} in the study of fourth order equations of the form $u_t = u_{xxxx} + G(u, u_x, u_{xx}, u_{xxx})$. And it was also considered by Castro Silva and Tenenblat \cite{silva2015} in the study of Camassa-Holm type equations of the form $u_t = u_{xxt} + \lambda u u_{xxx} + G(u, u_x, u_{xx}),(\lambda \in \mathbb{R})$, with quadratic nonlinear terms.

The comprehensive characterization results provided in \cite{chern1986, kamran1995,reyes1998} are highly valuable in determining whether a differential equation describes pseudospherical surfaces and in generating a wide range of such equations. For instance, the complete and explicit classifications for equations describing pseudospherical surfaces of type $u_t = A(x, t, u)u_{xx} + B(x, t, u, u_x)$ with $f_{21} = \eta$, $\eta \in \mathbb{R}$ and of type $u_{tt} = A(u, u_x, u_t)u_{xx} + B(u, u_x, u_t)u_{xt} + C(u, u_x, u_t)$ with  $f_{21,u_t} = f_{31,u_t} = 0$, were discussed in \cite{ferraioli2016,ferraioli2020}. Recently, Kelmer and Tenenblat  characterized systems of differential equations describing pseudospherical surfaces or spherical surfaces \cite{kelmer2022,kelmer2025}.

Around the time when the first draft of this work is finished, we learn that Catalano Ferraioli and Castro Silva have worked on the related problem \cite{ferraioli2024}, in which they consider the assumption that $f_{11}=\eta$ is constant and $f_{21}$ and $f_{31}$ are linear non-homogeneous functions of $u$ and $u_{xx}$ with constant coefficients. They arrived at three classification theorems by discussing the constant parameters in different situations. Moreover, they also researched the problem of determining sequences of conservation laws either in the pseudospherical or spherical cases. In this paper, a different auxiliary condition (\ref{1.4}) is selected, and thereby resulting in different equations. The generalized Camassa-Holm (gCH) equation is obtained therein, which is one of the well-known examples of CH-type equations, and its geometrical interpretations are provided as well. In addition, the nonexistence of equations of the form (\ref{1.1}) describing spherical surfaces can be readily deduced. It is worth highlighting that results in \cite{ferraioli2024} do not subsume our findings.

In this paper, we show families of equations contained in classification theorems. For examples,
\begin{flalign}\label{1.5}
 &(i)\quad  u_{t}-u_{xxt}=\lambda(u^{2}u_{xxx}-3u^{2}u_{x}-au^{3}+2u u_{x}u_{xx}+au^{2}u_{xx})-b(u-u_{xx})+\psi_{x}+a\psi, &
\end{flalign}
where $(\lambda a)^{2}+b^{2}\neq 0$ and $\psi(u,u_{x})$ is a differentiable function.

\noindent When $\lambda=1,a=1,b=0$ and $\psi=uu_{x}^{2}-2u^{2}u_{x}+u^{3}$, it is the gCH equation
\begin{equation}\label{1.6}
u_{t}-u_{xxt}=u^{2}u_{xxx}- u^2u_{xx} - 3u u_x^2 - 2u ^2u_x + 4u u_x u_{xx} + u_x^3, \end{equation}
which possesses an infinite hierarchy of local higher symmetries \cite{novikov2009}.
\begin{flalign}\label{1.7}
&(ii) \quad u_{t}-u_{xxt}=a\psi +u_x\psi_{u}+u_{xx}\psi_{u_x},\quad a\in\mathbb{R}/\{0\}, &
\end{flalign}
where $\psi(u,u_x)\neq0$ is a differentiable function.
\begin{flalign}\label{1.8}
& (iii) \quad u_{t}-u_{xxt}=u^{2}u_{xxx}-3u^2u_x+2uu_xu_{xx}-a(u_x^2+uu_{xx})-buu_x,  \quad a\neq0,b \in\mathbb{R}.&
\end{flalign}
\begin{flalign}\label{1.9}
& (iv)\quad  u_{t}-u_{xxt}=u^{2}u_{xxx}-5u^2u_x+4uu_xu_{xx}+(2\zeta_1-4)uu_x+2\zeta_1 u_x-2u_x u_{xx} ,\quad \zeta_1  \in\mathbb{R}.&
\end{flalign}

The remainder of this paper is organized as follows. In Section 2, we give a brief review in some basic results concerning the partial differential equations related to pseudospherical or spherical surfaces, in terms of linear problems associated to $\mathfrak{sl}(2, \mathbb{R})$-valued connection 1-form $ \omega_i $. The main results for classification will be shown in Section 3, Theorem 3.2-3.5, with explicit examples such as the gCH equation. The proofs of the theorems are presented in Section 4. And the last section is left for conclusion and discussion.

\section{Preliminaries}

\indent \indent In \cite{wadati1979}, Wadati et al. considered the following inverse scattering problem
\begin{equation}\label{2.1}
V_x=XV,\quad X=\left(
               \begin{array}{cc}
                 F(\eta) & H(\eta)q(x,t) \\
                 H(\eta)r(x,t) & -F(\eta) \\
               \end{array}
             \right),
\end{equation}
with $F(\eta)$ and $H(\eta)$ are functions of the eigenvalue $\eta$, and the time dependence of the eigenfunctions
\begin{equation}\label{2.2}
V_t=TV,\quad T=\left(
               \begin{array}{cc}
                 A(\eta,q,r) & B(\eta,q,r) \\
                 C(\eta,q,r) & -A(\eta,q,r) \\
               \end{array}
             \right),
\end{equation}
where $V = (v_1, v_2)^T$, with $v_i=v_i(x,t)$. By utilizing the compatibility condition of (\ref{2.1}) and (\ref{2.2}), namely $V_{xt}=V_{tx}$, and supposing that the eigenvalues $\eta$ are invariants, one can derive zero curvature representation \cite{crampin1977}
\begin{equation}\label{2.3}
X_t - T_x + [X, T] = 0,
\end{equation}
and the system for $A(\eta,q,r),B(\eta,q,r),C(\eta,q,r)$
\begin{equation}\label{2.4}
\begin{aligned}
  A_x+H(rB-qC)&=0,  \\
  Hq_t-B_x-2FB-2HqA&=0, \\
  Hr_t-C_x+2FC+2HrA&=0.
\end{aligned}
\end{equation}

Since exterior differential forms allow a compact and transparent presentation,  we will use them throughout. Then the inverse scattering problem (\ref{2.1}) and (\ref{2.2}) can be transformed into a completely integrable linear problem
\begin{equation}\label{2.5}
	dV = \Omega V,
\end{equation}
where $\Omega$ is a traceless $2\times 2$ matrix, provided by $\mathfrak{sl}(2, \mathbb{R})$-valued 1-form
\begin{equation}\label{2.6}
\Omega = \frac{1}{2} \begin{pmatrix} \omega_2 & \omega_1 - \omega_3 \\ \omega_1 + \omega_3 & -\omega_2 \end{pmatrix},
\end{equation}
with associated 1-forms $\omega_i$ defined by
\begin{equation}\label{2.7}
\begin{aligned}
  \omega_1&=H(r+q)\,dx+(B+C)\,dt,  \\
  \omega_2&=2F \,dx+2A\,dt,\\
  \omega_3&=H(r-q)\,dx+(C-B)\,dt.
\end{aligned}
\end{equation}
Note that integrability of (\ref{2.5}) is
\begin{equation}\label{2.8}
	d\,\Omega - \Omega \wedge \Omega = 0,
\end{equation}
which can be verified to be equivalent to the following relations
\begin{equation}\label{2.9}
d\omega_1 = \omega_3 \wedge \omega_2, \quad d\omega_2 = \omega_1 \wedge \omega_3,\quad d\omega_3 = \omega_1 \wedge \omega_2.
\end{equation}
In the pursuit of solving (\ref{2.4}) or (\ref{2.8}) for $F,H,A,B,C$, it is commonly ascertained to satisfy another partial differential equation.

Let $S$ be a two dimensional Riemannian manifold equipped with a coframe $\{\omega_1,\omega_2\}$ dual to an orthogonal frame $\{e_1, e_2\}$. The metric on $S$ can then be expressed as $g = \omega_1^2 + \omega_2^2$. The first two equations in (\ref{2.9}) are the structure equations determining the connection form $\omega_3:=\omega_{12}$, while the last equation in (\ref{2.9}), known as the Gauss equation, implies that the Gaussian curvature of $S$ is -1, i.e. $S$ is a pseudospherical surface. Furthermore, the existence of 1-forms (\ref{2.7}) satisfying (\ref{2.9}) requires that a certain partial differential equation must be satisfied.

A $k$-th order differential equation for a real-valued function $u(x, t)$ is said to describe pseudospherical surfaces ($\delta = 1$) (or spherical surfaces ($\delta = -1$)) if there exist 1-forms $\omega_i = f_{i1} dx + f_{i2} dt$, $1 \leq i \leq 3$, with $\omega_1 \wedge \omega_2 \neq 0$, where the coefficient functions $f_{ij}$, $j = 1, 2$, depend on $u(x, t)$ and its derivatives, such that the structure equations of a surface with Gaussian curvature $K = -1$ (or $K = 1$), say
\begin{equation}\label{2.10}
	d\omega_1 = \omega_3 \wedge \omega_2, \quad d\omega_2 = \omega_1 \wedge \omega_3, \quad d\omega_3 = \delta \omega_1 \wedge \omega_2,
\end{equation}
are satisfied whenever $u(x,t)$ is a solution of the differential equation.

It is essential to observe that the wedge product $\omega_1 \wedge \omega_2$ is non-zero for solutions $u(x, t)$ to equations describing pseudospherical or spherical surfaces. However, this condition does not imply that $\omega_1 \wedge \omega_2$ is non-zero at every point in the domain $U \subseteq \mathbb{R}^2$. For generic solutions, $\omega_1 \wedge \omega_2$ is non-zero almost everywhere in $U$, meaning it can vanish only on a set of measure zero. For any such generic solution $u(x, t)$, the metric $g = \omega_1^2 + \omega_2^2$ defines a Riemannian metric on $U$ with Gaussian curvature $K = -\delta $ almost everywhere, which is an intrinsic geometric property (not immersed into any ambient space).

\begin{Remark}
The $2\times 2$ matrix $\Omega$ is not unique for linear problem (\ref{2.5})  and the integrability condition (\ref{2.8}) is invariant under the gauge transformation \cite{sasaki1979}
\begin{equation}\label{2.11}
\Omega \rightarrow \Omega^{\prime}=dAA^{-1}+A\Omega A^{-1},
\end{equation}
with $A$ a $2\times2$ matrix satisfying $\det A=1$. In fact, choosing
\begin{equation}\label{2.12}
A=\frac{\sqrt{2}}{2} \begin{pmatrix}-i & 1 \\ 1 & -i\end{pmatrix},
\end{equation}
we obtain $\mathfrak{su}(2)$-valued 1-form
\begin{equation}\label{2.13}
\Omega = \frac{1}{2} \begin{pmatrix} i\omega_3 & \omega_1 -i\omega_2 \\ \omega_1 + i\omega_2 & -i\omega_3 \end{pmatrix}.
\end{equation}
\end{Remark}

\begin{Remark}
Given that the local isomorphism between $SL(2,\mathbb{R})$ and $SO(2,1)$ provides  the Lie algebra isomorphism $\mathfrak{so}(2,1)\cong\mathfrak{sl}(2,\mathbb{R})$, and the local isomorphism between $SO(3)$ and $SU(2)$ provides the Lie algebra isomorphism $\mathfrak{so}(3)\cong\mathfrak{su}(2)$. Hence, we can find $\mathfrak{so}(2,1)$ (resp. $\mathfrak{so}(3)$)-valued 1-form \cite{kelmer2022}
\begin{equation}\label{2.14}
\widetilde{\Omega}= \begin{pmatrix} 0 & \omega_1 & \omega_2 \\ \delta \omega_1 & 0 & \omega_3 \\ \delta \omega_2 & -\omega_3 & 0 \end{pmatrix},
\end{equation}
with $\delta = 1$ (resp. $\delta = -1$).
\end{Remark}

\begin{Remark}
An equation possessing a zero curvature representation does not necessarily imply that the equation can describe pseudospherical or spherical surfaces. For an accurate description of pseudospherical or spherical surfaces, the coefficient functions $f_{ij}$ in (\ref{2.10}) must satisfy the non-degeneracy condition $$ f_{11} f_{22} - f_{12} f_{21} \neq 0.$$
\end{Remark}

As of now, there are a large number of equations describing pseudospherical of spherical surface.  The most representative example is the SG equation \cite{chern1986}
\begin{equation}\label{2.15}
u_{xt}=\sin u,
\end{equation}
with associated 1-forms
\begin{equation}\label{2.16}
\omega_1=\frac{1}{\eta}\sin u\,dt,\quad \omega_2=\eta\,dx+\frac{1}{\eta}\cos u\,dt,\quad \omega_3=u_x\,dx,
\end{equation}
where $\eta\in\mathbb{R}/\{0\}$. Also, the classical dispersive wave equation which describes pseudospherical surfaces is the Degasperis-Procesi (DP) equation \cite{silva2015}
\begin{equation}\label{2.17}
u_t - u_{xxt} =u u_{xxx} + 3u_{x} u_{xx} - 4u u_x,
\end{equation}
with associated 1-forms
\begin{equation}\label{2.18}
\begin{aligned}
\omega_1&=(u-u_{xx})\,dx+(uu_{xx}-2uu_x+u_x^2)\,dt, \\
\omega_2&=\left(\mu(u-u_{xx})\pm2\sqrt{1+\mu^2}\right)\,dx+\mu(uu_{xx}-2uu_x+u_x^2)\,dt, \\
\omega_3&=\left(\pm\sqrt{1+\mu^2}(u-u_{xx})+2\mu\right)\,dx\pm\sqrt{1+\mu^2}(uu_{xx}-2uu_x+u_x^2)\,dt, \\
\end{aligned}
\end{equation}
where $\mu\in\mathbb{R}$. Indeed, for the 1-forms above, the structure equations (\ref{2.10}), with $\delta=1$, are satisfied modulo (\ref{2.17}).

On the other hand, a celebrated instance of equation describing spherical surfaces is the $NLS^{+}$ equation \cite{ding2002}
\begin{equation}\label{2.19}
  \begin{cases}
  u_t+v_{xx}+2(u^2+v^2)u=0,\\
  -v_t+u_{xx}+2(u^2+v^2)v=0,
  \end{cases}
\end{equation}
with associated 1-forms
\begin{equation}\label{2.20}
\begin{aligned}
\omega_1&=2v\,dx+2(-2\eta v+u_x)\,dt, \\
\omega_2&=2\eta\,dx+2(-2\eta^2+u^2+v^2)\,dt, \\
\omega_3&=-2u\,dx+2(\eta u+v_x)\,dt, \\
\end{aligned}
\end{equation}
where $\eta\in\mathbb{R}$. Indeed, for the 1-forms above, the structure equations (\ref{2.10}), with $\delta=-1$, are satisfied modulo (\ref{2.19}).

Hereafter, for the convenience of discussion, we adopt the notation,
\begin{equation}\label{2.21}
u_1 = \frac{\partial u}{\partial x}, \quad u_2 = \frac{\partial ^{2}u}{\partial x^2}, \quad u_3 = \frac{\partial ^{3}u}{\partial x^3}, \quad \cdots, \quad u_k=\frac{\partial ^{k}u}{\partial x^k}.
\end{equation}

\section{Main result and special classes of equations}
\indent \indent The present section will concentrate on the classification of third order equations of the form
\begin{equation}\label{3.1}
	u_{t} - u_{2,t} = \lambda u^2 u_{3} + G(u, u_{1}, u_{2}), \quad \lambda\in\mathbb{R}, \quad G \neq 0,
\end{equation}
which describes pseudospherical or spherical surfaces. To this end, we assume the auxiliary condition for the coefficient functions
\begin{equation}\label{3.2}
f_{p1} = \mu_p f_{11} + \eta_p, \quad \mu_p, \eta_p \in \mathbb{R}, \quad p=2,3.
\end{equation}
The main results for the classification are summarized in Theorem 3.2-3.5, successively. Theorem 3.2-3.5 are based on Lemma 3.1 below, which states the existence conditions for the coefficient functions $f_{ij}$ and $G$ guaranteing the corresponding equation that can describe pseudospherical or spherical surfaces. The complete proofs of these lemma and theorems will appear in the next section.

\begin{Lemma}
 A partial differential equation of the form (\ref{3.1}) with associated 1-forms $\omega_i=f_{i1}dx+f_{i2}dt, 1\leq i\leq3$, where the coefficient functions $f_{ij}=f_{ij}(u,u_1,\ldots,u_k)$ satisfying the condition (\ref{3.2}), describes pseudospherical surfaces ($\delta=1$) or spherical surfaces ($\delta=-1$) if and only if $f_{ij}$ and $G$ satisfy the following conditions:
\begin{equation}\label{3.3}
f_{i1,u_1}=0,\quad f_{i1,u_k}=f_{i2,u_k}=0,\quad 3\leq k \leq m,\quad m\in\mathbb{Z},
\end{equation}
\begin{equation}\label{3.4}
f_{i1,u}+f_{i1,u_2}=0,
\end{equation}
\begin{equation}\label{3.5}
f_{i2}=-\lambda u^{2}f_{i1}+\phi_{i},
\end{equation}
where $\phi_{i}=\phi_{i}(u,u_{1})$ are real differential functions of $u,u_{1}$ satisfying
\begin{equation}\label{3.6}
-Gf_{11,u}+(-2\lambda u f_{11}-\lambda u^2 f_{11,u}+\phi_{1,u})u_{1}+\phi_{1,u_1}u_{2}+Mf_{11}+N=0,
\end{equation}
\begin{equation}\label{3.7}
Qf_{11}+L_{2,u}u_{1}+L_{2,u_1}u_{2}-2\lambda \eta_{2}u u_{1}-\mu_{2}N+\eta_{3}\phi_{1}=0,
\end{equation}
\begin{equation}\label{3.8}
-(\delta L_2+\mu_{3}M)f_{11}+L_{3,u}u_{1}+L_{3,u_1}u_{2}-2\lambda \eta_{3}u u_{1}-\mu_{3}N+\delta\eta_{2}\phi_{1}=0,
\end{equation}
\begin{equation}\label{3.9}
-L_2 f_{11}+\eta_{2}\phi_{1}\neq 0,
\end{equation}
with
\begin{equation}\label{3.10}
\begin{split}
 & L_p= L_p(u, u_1) := \phi_{p} - \mu_p\phi_{1},\quad p=2,3, \\
 & M=M(u,u_1):=\mu_{2}\phi_{3}-\mu_{3}\phi_{2}, \\
 & N=N(u,u_1):=\eta_{2}\phi_{3}-\eta_{3}\phi_{2}, \\
 & Q = Q(u, u_1) := -(L_3+\mu_2 M), \\
 & \gamma := \mu_2\mu_3\eta_2 - (1 + \mu_2^2)\eta_3.
\end{split}
\end{equation}
\end{Lemma}

\subsection*{3.1 Classification Theorems}

\indent \indent  In this subsection, we summarize the main results for classification. For each scenario, we will ascertain the functions $\phi_{i}$ as stipulated in (\ref{3.5}).

\begin{Theorem}
Consider a partial differential equation of the form (\ref{3.1}) which describes pseudospherical surfaces or spherical surfaces, with coefficient functions $f_{ij}=f_{ij}(u,u_1,\ldots, u_k)$ satisfying (\ref{3.2})-(\ref{3.9}) with $Q =L_2= 0,\gamma=0$, if and only if $\delta=1$ and the equation (\ref{3.1}) can be written in the form
\begin{equation*}
u_t-u_{2,t}=\frac{1}{f'}\left(\phi_{1,u}u_{1}+\phi_{1,u_1}u_{2}\pm \frac{\eta_2}{\sqrt{1+\mu_{2}^2}}\phi_{1}\right),\quad \eta_2\neq0.
\end{equation*}
Moreover, associated 1-forms are
\begin{align*}
&\omega_{1}=f\,dx+\phi_{1}\,dt,
\\&\omega_{2}=(\mu_{2}f_{11}+\eta_{2})\,dx+\mu_{2}\phi_{1}\,dt,
\\&\omega_{3}=\pm \left(\sqrt{1+\mu_{2}^2}f_{11}+\frac{\mu_{2} \eta_2}{\sqrt{1+\mu_{2}^2}}\right)\,dx\pm \sqrt{1+\mu_{2}^2}\phi_{1}\,dt,
\end{align*}
where $\phi_{1} \neq 0$ and $f = f(u - u_2)$ is a differentiable function such that $f' \neq 0$.
\end{Theorem}

\begin{Theorem}
Consider a partial differential equation of the form (\ref{3.1}) which describes pseudospherical surfaces or spherical surfaces, with coefficient functions $f_{ij}=f_{ij}(u,u_1,\ldots, u_k)$ satisfying (\ref{3.2})-(\ref{3.9}) with $Q =L_2=0,\gamma\neq 0$, if and only if $\delta=1$ and the equation (\ref{3.1}) can be written in the form
\begin{equation*}
u_t-u_{2,t}= \lambda u^2 u_{3}
-\frac{\lambda}{f'}\left(2u u_{1}f+u^{2}u_{1}f'+\frac{2\eta_{2}}{\gamma}(u_{1}^{2}+u u_{2}+(\mu_{3}\eta_{2}-\mu_{2}\eta_{3})u u_{1})\right),\quad \lambda \eta_2\neq0.
\end{equation*}
Moreover, associated 1-forms are
\begin{align*}
&\omega_{1}=f\,dx-\lambda\left( u^{2}f+\frac{2}{\gamma}\eta_{2} u u_{1}\right)\,dt,
\\&\omega_{2}=(\mu_{2}f+\eta_{2})\,dx-\lambda \left(u^{2}f_{21}+\frac{2}{\gamma}\mu_{2}\eta_{2} u u_{1}\right)\,dt,
\\&\omega_{3}=(\mu_{3}f+\eta_{3})\,dx-\lambda \left(u^{2}f_{31}+\frac{2}{\gamma}\mu_{3}\eta_{2} u u_{1}\right)\,dt,
\end{align*}
where $\eta_{2}^{2}-\eta_{3}^{2}-(\mu_{2}\eta_{3}-\mu_{3}\eta_{2})^{2}=0$ and $f=f(u- u_2)$ is a differentiable function with $f'\neq0$.
\end{Theorem}	

\begin{Theorem}
Consider a partial differential equation of the form (\ref{3.1}) which describes pseudospherical surfaces or spherical surfaces, with coefficient functions $f_{ij}=f_{ij}(u,u_1,\ldots, u_k)$ satisfying (\ref{3.2})-(\ref{3.9}) with $Q =0, L_2\neq 0,\gamma=0$, if and only if $\delta=1$ and the equation (\ref{3.1}) can be written in the form
\begin{equation*}
\begin{split}
u_t-u_{2,t}= &\lambda u^2 u_{3}+\frac{1}{f'}\left(u_{1}\phi_{1,u}+u_{2}\phi_{1,u_1}-\lambda u^{2}u_{1}f'\pm\frac{\eta_2}{\sqrt{1+\mu_{2}^2}}\phi_{1}\right.
\\&\left.-\left(2\lambda u u_{1}\pm\frac { \eta_2 }{\sqrt{1+\mu_{2}^2}}\lambda u^{2}\pm\frac{C_1}{\sqrt{1+\mu_{2}^2}}\right)f\right), \quad C_1
\in \mathbb{R}.
\end{split}
\end{equation*}
Moreover, associated 1-forms are
\begin{align*}
&\omega_{1}=f\,dx-(\lambda u^{2}f-\phi_{1})\,dt,
\\&\omega_{2}=(\mu_{2}f+\eta_{2})\,dx-(\lambda \mu_{2}u^{2}f-\mu_{2} \phi_{1}-C_1)\,dt,
\\&\omega_{3}=\pm \left(\sqrt{1+\mu_{2}^2}f+ \frac{\mu_{2} \eta_2}{\sqrt{1+\mu_{2}^2}}\right)\,dx\mp \sqrt{1+\mu_{2}^2}\left(\lambda u^{2}f- \phi_{1}- \frac{\mu_{2}C_1}{1+\mu_{2}^2}\right)\,dt,
\end{align*}
where $(\lambda \eta_2)^{2}+C_1^{2} \neq 0$ and $f=f(u-u_2)$ is a differentiable function such that $f'\neq 0$.
\end{Theorem}

\begin{Theorem}
Consider a partial differential equation of the form (\ref{3.1}) which describes pseudospherical surfaces or spherical surfaces, with coefficient functions $f_{ij}=f_{ij}(u,u_1,\ldots, u_k)$ satisfying (\ref{3.2})-(\ref{3.9}) with $Q\neq0 ,L_2\neq 0,\gamma\neq0$, if and only if $\delta=1$ and the equation (\ref{3.1}) can be written in one of the following two forms,
\\(i)
\begin{align*}
u_t-u_{2,t}=& \lambda u^2 u_{3}+\lambda \left(-5u^{2}u_{1}+4u u_{1}u_{2}+\left(2\zeta_{1}-\frac{4}{\theta}\right)u u_{1}
-\frac{2}{\theta}u_{1}u_{2}+\frac{2\zeta_{1}}{\theta}u_{1}\right)
\\&+\left(\theta u_{1}^{3}+2u u_{1}+u_{1}u_{2}-\zeta_{1}u_{1}\right)\theta C_2 e^{\theta u},\quad 0\neq\nu\theta,\sigma,C_2\in\mathbb{R},\quad \lambda^{2}+C_2^{2}\neq 0,
\end{align*}
and associated 1-forms are
\begin{align*}
\omega_{1}=&[\nu(u-u_2)-\sigma]\,dx-\left(\lambda u^{2}f_{11}+\frac{\nu}{\theta}\left(2\lambda -\theta^{2}C_2 e^{\theta u}\right)u_{1}^{2}\right.\\
&+\left.\left(\frac{2\lambda}{\theta}-\theta C_2 e^{\theta u}+2\lambda u\right)
\left(\frac{\nu u-\sigma}{\theta}\pm\left(\mu_{2}-\frac{\nu\eta_{2}}{\theta}\right)\frac{u_1}{\sqrt{1+\mu_{2}^2}}\right)\right)\,dt,\\
\omega_{2}=&(\mu_{2}f_{11}+\eta_{2})\,dx+\left(\mu_{2}f_{12}-\lambda \eta_{2}u^{2}
+\left(\frac{2\lambda}{\theta}-\theta C_2 e^{\theta u}+2\lambda u\right)\left(\pm \sqrt{1+\mu_{2}^2}u_{1}-\frac{\eta_{2}}{\theta}\right)\right)\,dt,\\
\omega_{3}=&\pm \left(\sqrt{1+\mu_{2}^2}f_{11}\pm\eta_3\right)\,dx \\
& \pm\left( \sqrt{1+\mu_{2}^2}f_{12}\mp\lambda \eta_{3}u^{2}\pm\left(\frac{2\lambda}{\theta}-\theta C_2 e^{\theta u}+2\lambda u \right)
\left(\mu_{2}u_{1}-\eta_3\right)\right)\,dt,
\end{align*}
with $$\zeta_{1}=\frac{2\sigma}{\nu}-\frac{1}{\theta}-\frac{\theta}{\nu^{2}(1+\mu_{2}^{2})}-\frac{\eta_{2}(2\theta \mu_{2}+\nu\eta_{2})}{\theta \nu(1+\mu_{2}^{2})}.$$
\\(ii)
\begin{align*} u_t-u_{2,t}  = &  \lambda u^2 u_{3}+\lambda \left(-3u^{2}u_{1}+2u u_{1}u_{2}+2\zeta_{2}u u_{1}\mp \frac{2}{\tau}(u_{1}^{2}+u u_{2})\right)+\varphi^{''}u_{1}^{2}e^{\pm \tau u_1}\\
& +\tau \left(\tau u u_{2}\pm u_{1}-\zeta_{2}\tau u_{2}\right)\varphi e^{\pm \tau u_1}
\pm(\tau u u_{1}+\tau u_{1}u_{2}\pm u_{2}-\zeta_{2}\tau u_{1})\varphi^{'}e^{\pm \tau u_1},
\end{align*}
with  $0<\tau,\nu,\sigma\in \mathbb{R}, \nu\eta_2\neq0$. And the associated 1-forms are
\begin{align*}
\omega_{1}=&(\nu(u-u_2)-\sigma)\,dx
-\left(\lambda u^{2}f_{11}-(\pm\tau(\nu u-\sigma)\varphi+\nu\varphi 'u_{1})e^{\pm \tau u_1}\pm \frac{2\lambda \nu}{\tau}u u_{1}\right)\,dt,\\
\omega_{2}=&(\mu_{2}f_{11}+\eta_{2})\,dx+
\left(\mu_{2}f_{12}-\lambda \eta_{2}u^{2}\pm \tau \eta_{2}\varphi e^{\pm \tau u_1}\right)\,dt,\\
\omega_{3}=&\pm\tau \left(\left(\frac{\sigma}{\nu}-\zeta_{2}\right)\left(\frac{1+\mu_{2}^{2}}{\eta_2}f_{11}+\mu_{2}\right)- \frac{1}{\nu}f_{21}\right)\,dx\\
&\pm\tau\left( \left(\frac{\sigma}{\nu}-\zeta_{2}\right)\left(\frac{1+\mu_{2}^{2}}{\eta_2}f_{12}-\mu_{2}\left(\lambda u^{2}\mp\tau \varphi e^{\pm\tau u_1}\right)\right)- \frac{1}{\nu}f_{22}\right)\,dt,
\end{align*}
where $\varphi(u) \neq 0$ is arbitrary differentiable function and $$\zeta_{2}=\frac{\sigma}{\nu}\mp\frac{\mu_{3}\eta_{2}-\mu_{2}\eta_{3}}{\tau}.$$
\end{Theorem}

\begin{Corollary}
There are no partial differential equations of the form (\ref{3.1}) that can describe spherical surfaces with associated 1-forms $\omega_i=f_{i1}dx+f_{i2}dt, 1\leq i\leq3$, where the coefficient functions $f_{ij}=f_{ij}(u,u_1,\ldots,u_k)$ satisfying the condition (\ref{3.2}).
\end{Corollary}

\subsection*{3.2 Special classes of equations}
	
\indent \indent In this subsection, examples of equations derived from applying classification Theorems 3.2 to 3.5 are presented.

\begin{Example} In Theorem 3.4, let $ f = u - u_2 $, then
\begin{equation}\label{3.11}
u_{t}-u_{2,t}=\lambda(u^{2}u_{3}-3u^{2}u_{1}-au^{3}+2u u_{1}u_{2}+au^{2}u_{2})-b(u-u_{2})+u_{1}\psi_{u}+u_{2}\psi_{u_{1}}+a\psi,
\end{equation}
where $(\lambda a)^{2}+b^{2}\neq 0$ and $\psi(u,u_{1})$ is a differentiable function. Furthermore, by taking $\lambda=1,a=1,b=0$ and $\psi(u,u_{1})=u u_{1}^{2}-2u^{2}u_{1}+u^{3}$, we get the nonlinear equation
\begin{equation}\label{3.12}
u_{t} - u_{2,t} = u^2 u_3 - u^2u_2 - 3u u_1^2 - 2u ^2u_1 + 4u u_1u_2 + u_1^3,
\end{equation}
which is the gCH equation proposed by Novikov in \cite{novikov2009}, and associated 1-forms
\begin{equation}\label{3.13}
\begin{aligned}
\omega_{1} &= (u - u_2)\,dx+( u^2 u_2 - 2u^2 u_1 + u u_1^2)\,dt, \\
\omega_{2} &=\left( \mu(u- u_2) \pm \sqrt{1 + \mu^2}\right)\,dx+ \mu(u^2 u_2 - 2u^2 u_1 + u u_1^2)\,dt, \\
\omega_{3} &= \left(\pm\sqrt{1 + \mu^2}(u - u_2) + \mu\right)\,dx\pm\sqrt{1 + \mu^2}(u^2 u_2 - 2u^2 u_1 + u u_1^2)\,dt,
\end{aligned}
\end{equation}
with $\mu\in\mathbb{R}$. The corresponding linear problem (\ref{2.5}) is dependent on the parameter $\mu$.

Here we articulate the geometrical interpretation of the gCH equation (\ref{3.12}). Let $S$ be a surface with Gaussian curvature $K= -1$ (i.e., pseudospherical surface) and its parametric equation is $\mathbf{r=r}(x,t)$. Consider an orthonormal coordinate grid on $S$, and let $\mathbf{e}_1$, $\mathbf{e}_2$ be the unit tangent vectors, and $\mathbf{e}_3$ be the unit normal vector. Then the coframes $\omega_i$ and $\omega_{ij}$ satisfy
\begin{equation}\label{3.14}
\begin{cases}
d\mathbf{r} = \omega_1 \mathbf{e}_1 + \omega_2 \mathbf{e}_2, \quad \omega_3 = 0, \\
d\mathbf{e}_1 = \omega_{12} \mathbf{e}_2 + \omega_{13} \mathbf{e}_3, \\
d\mathbf{e}_2 = \omega_{21} \mathbf{e}_1 + \omega_{23} \mathbf{e}_3, \\
d\mathbf{e}_3 = \omega_{31} \mathbf{e}_1 + \omega_{32} \mathbf{e}_2,
\end{cases}
\end{equation}
with $\omega_{ij}+\omega_{ji}=0$, $i,j=1,2,3$.  When $\omega_1, \omega_2$ and $\omega_3:=\omega_{12}$ is taken as in (\ref{3.13}), we conclude that the metric of the surface $S$ is
\begin{equation}\label{3.15}
\begin{aligned}
g=&\left((u-u_2)^2+\left( \mu(u- u_2) \pm \sqrt{1 + \mu^2}\right)^2\right)\,dx^2 \\
&+2\left((1+\mu^2)(u-u_2)\pm\mu \sqrt{1 + \mu^2}\right)( u^2 u_2 - 2u^2 u_1 + u u_1^2)\,dxdt \\
&+(1+\mu^2)( u^2 u_2 - 2u^2 u_1 + u u_1^2)^2\,dt^2,
\end{aligned}
\end{equation}
and the structure equations
\begin{equation}\label{3.16}
d\omega_1 = \omega_3 \wedge \omega_2, \quad d\omega_2 = \omega_1 \wedge \omega_3,\quad d\omega_3 = \omega_1 \wedge \omega_2,
\end{equation}
satisfy if and only if (\ref{3.12}) holds. In other words, on a pseudospherical surface, there exists parameters $(x, t)$ such that its structure equations transform into the gCH equation.

 Conversely, given any solution of the gCH equation, we compute associated 1-forms ${\omega_i,\omega_{ij}}$. Since the structure equations hold, by substituting 1-forms into (\ref{3.14}), we can construct a surface with Gaussian curvature equal to -1
 \begin{equation}\label{3.17}
 S:\mathbf{r}=\mathbf{r}(x,t).
 \end{equation}
\end{Example}

\begin{Example}
Theorem 3.2 gives a family of equations describing pseudospherical surfaces
\begin{equation}\label{3.18}
u_{t}-u_{2,t}=a\psi+u_{1}\psi_{u}+u_{2}\psi_{u_1}, \quad a\in\mathbb{R}/ \{0\},
\end{equation}
where $\psi(u,u_{1})\neq0$ is a differentiable function. In fact, considering $f=u-u_{2}$ and $a=\pm\eta_{2}/\sqrt{1+\mu_{2}^{2}}$ in Theorem 3.2, we have associated 1-forms
\begin{equation}\label{3.19}
\begin{aligned}
\omega_{1}&=(u-u_{2})\,dx+\psi\,dt,\\
\omega_{2}&=\left(\mu(u-u_{2})\pm \sqrt{1+\mu_{2}^{2}}a\right)\,dx+\mu \psi\,dt,\\
\omega_{3}&=\left(\pm\sqrt{1+\mu^{2}}(u-u_{2})+\mu a\right)\,dx\pm\sqrt{1+\mu^{2}}\psi\,dt,
\end{aligned}
\end{equation}
with $\mu\in\mathbb{R}$. The corresponding linear problem (\ref{2.5}) depends on the parameter $ \mu$. For instance,

\noindent (i) if $ \psi(u,u_1)= u u_1 $, then we get the equation
\begin{equation}\label{3.20}
u_{t}-u_{2,t}=a u u_{1}+u_{1}^{2}+u u_{2}.
\end{equation}
(ii) if $ \psi(u,u_1) = e^{u u_1} $, then we get the equation
\begin{equation}\label{3.21}
u_{t}-u_{2,t}=ae^{u u_{1}}+(u_{1}^{2}+u u_{2})e^{u u_{1}}.
\end{equation}
(iii) if $\psi(u,u_1)=u^{p}$, then we have
\begin{equation}\label{3.22}
u_{t}-u_{2,t}=au ^{p}+pu^{p-1}u_{1},\quad p\in\mathbb{Z}/\{0\}.
\end{equation}
(iv) if $\psi(u,u_1)=u_{1}^{p}$, then we have
\begin{equation}\label{3.23}
u_{t}-u_{2,t}=au_{1}^{p}+pu_{1}^{p-1}u_{2},\quad p\in\mathbb{Z}/\{0\}.
\end{equation}
\end{Example}

\begin{Example}
In Theorem 3.3, without loss of generality, setting $ \lambda = 1 $, there yields a family of equations that describe pseudospherical surfaces
\begin{equation}\label{3.24}
u_t-u_{2,t}=u^2 u_{3}
-\frac{1}{f'}\left(2u u_{1}f+u^{2}u_{1}f'+a(u_{1}^{2}+u u_{2})+bu u_{1}\right),\quad 0\neq a, b\in\mathbb{R}.
\end{equation}
Note that if $b\neq 0$, the parameter $\eta$ must satisfy $|\eta|>|b|/|a|$ and $\mu$ is given in terms of $\eta$, $a$ and $b$. On the other hand, if $b=0$, one concludes that $a=\pm2$.

\noindent (i) if $ f = u - u_2$, then one finds the equation
\begin{equation}\label{3.25}
u_{t}-u_{2,t}=u^{2}u_{3}-3u^{2}u_{1}+2u u_{1}u_{2}\mp2(u_{1}^{2}+u u_{2}),
\end{equation}
and associated 1-forms
\begin{equation}\label{3.26}
\begin{aligned}
\omega_{1}&=(u-u_{2})\,dx-(u^{2}(u-u_{2})\pm2u u_{1})\,dt,\\
\omega_{2}&=(\mu(u-u_{2})+\eta)\,dx-\left(\mu u^{2}(u-u_{2})+\eta u^{2}\pm 2\mu u u_{1}\right)\,dt,\\
\omega_{3}&=\mp(\mu(u-u_{2})+\eta)\,dx\pm(\mu u^{2}(u-u_{2})+\eta u^{2}\pm 2\mu u u_{1})\,dt,
\end{aligned}
\end{equation}
with $\mu,\eta\in\mathbb{R}$ and $\eta\neq0$. The corresponding linear problem (\ref{2.5}) depends on the parameters $\mu$ and $\eta$.

\noindent (ii) if $f=e^{u-u_{2}}$, then one finds the equation
\begin{equation}\label{3.27}
u_{t}-u_{2,t}=u^{2}u_{3}-2u u_{1}-u^{2}u_{1}\mp 2(u_{1}^{2}+u u_{2})e^{u_{2}-u}.
\end{equation}
(iii) if $f=(u-u_{2})^{p}$, then one finds the equation
\begin{equation}\label{3.28}
u_{t}-u_{2,t}=u^{2}u_{3}-u^{2}u_{1}-\frac{1}{p}[2u u_{1}\pm 2(u_{1}^{2}+u u_{2})(u-u_{2})^{-p}](u-u_{2}),\quad p\in\mathbb{Z}/\{0\}.
\end{equation}
\end{Example}

\begin{Example} Let $\theta=1$, $C_2=0$, and $\lambda=1$ for simplicity in Theorem 3.5(i), we obtain a family of equations describing pseudospherical surfaces
\begin{equation}\label{3.29}
u_{t}-u_{2,t}=u^{2}u_{3}-5u^{2}u_{1}+4u u_{1}u_{2}+(2\zeta_{1}-4)u u_{1}+2\zeta_1u_{1}-2 u_{1}u_{2},
\end{equation}
in terms of 4 parameters namely, $\nu$, $\sigma$, $\mu$ and $\eta$. Opting $\mu=0$ and $\nu=1$ yields associated 1-forms
\begin{equation}\label{3.30}
\begin{aligned}
\omega_{1}&=(u-u_{2}-\sigma)\,dx-[u^{2}f_{11}+2u_{1}^{2}+(2u+2)(u-\sigma\mp\eta u_{1})]\,dt,\\
\omega_{2}&=\eta\,dx-[\eta u^{2} -(2u+2)(\pm u_{1}-\eta)]\,dt,\\
\omega_{3}&=\pm\left(u-u_{2}-\frac{\zeta_{1}+\eta^{2}}{2}\right)\,dx\pm [f_{12}-u^{2}-(2u+2)]\,dt,
\end{aligned}
\end{equation}
with $\sigma=1+(\zeta_{1}+\eta^{2})/2$, $\zeta_1$ defined as (\ref{4.57}) and $\eta\in\mathbb{R}/\{0\}$. The corresponding linear problem (\ref{2.5}) depends on the parameter $\eta$.
\end{Example}
	
\begin{Example} Theorem 3.5(ii) formulates the equations using a non-zero real differential function $ \varphi(u) $. Setting $\varphi(u)=1/\tau$, we see a family of equations describing pseudospherical surfaces
\begin{equation}\label{3.31}
\begin{split}
u_{t}-u_{2,t}=&\lambda\left(u^{2}u_{3}-3u^{2}u_{1}+2u u_{1}u_{2}+2\zeta_{2}u u_{1}\mp\frac{2}{\tau}(u_{1}^2+u u_{2})\right)\\
&+(\tau u u_{2}\pm u_{1}-\zeta_{2}\tau u_{2}) e^{\pm\tau u_1},
\end{split}
\end{equation}
and associated 1-forms
\begin{equation}\label{3.32}
\begin{split}
\omega_{1}=&(\nu(u-u_2)-\sigma)\,dx-\left(\lambda u^{2}f_{11}\mp(\nu u-\sigma)e^{\pm \tau u_1}\pm \frac{2\lambda \nu}{\tau}u u_{1}\right)\,dt,\\
\omega_{2}=&(\mu f_{11}+\eta)\,dx+\left(\mu f_{12}-\lambda \eta u^{2}\pm \eta  e^{\pm \tau u_1}\right)\,dt,\\
\omega_{3}=&\pm\left(\tau \left(\frac{\sigma}{\nu}-\zeta_{2}\right)\left(\frac{1+\mu^{2}}{\eta}f_{11}+\mu\right)- \frac{\tau}{\nu}f_{21}\right)\,dx\\
&\pm\left(\tau \left(\frac{\sigma}{\nu}-\zeta_{2}\right)\left(\frac{1+\mu^{2}}{\eta}f_{12}-\mu\left(\lambda u^{2}\mp e^{\pm\tau u_1}\right)\right)- \frac{\tau}{\nu}f_{22}\right)\,dt,
\end{split}
\end{equation}
with $\mu,\eta\in\mathbb{R}$ and $\eta\neq0$. The corresponding linear problem (\ref{2.5}) depends on the parameters $\mu$ and $\eta$.
\end{Example}

\begin{Remark}
The Novikov equation \cite{novikov2009}
\begin{equation}\label{3.33}
u_{t} - u_{2,t} = u^2u_3 - 4u^2 u_1+ 3u u_1u_2,
\end{equation}
does not satisfy the applicable classification theorems. Therefore, under the condition (\ref{3.2}), the Novikov equation does not describe pseudospherical surfaces.
\end{Remark}

\section{Proof of the main results}
\noindent\textbf{4.1 Proof of  Lemma 3.1}
\begin{Proof}
Let $u(x,t)$ be a function satisfying (\ref{3.1}). Then
\begin{equation}\label{4.1}
du\wedge dx = (\lambda u^2 u_3 + G)\,dt \wedge dx + du_2 \wedge dx,
\end{equation}
and for $0 \leq k \leq m - 1$,
\begin{equation}\label{4.2}
du_k \wedge dt = u_{k+1}\,dx \wedge dt.
\end{equation}
Since the functions $f_{ij}$ are independent of $x$ and $t$ explicitly, it follows that
\begin{equation}\label{4.3}
df_{ij} = \sum_{k=0}^{m} f_{ij,u_k}\,du_k.
\end{equation}
Applying (\ref{3.1}), we obtain
\begin{equation}\label{4.4}
 \begin{split}
d \omega_{i}=&\left(-(\lambda u^{2}u_{3}+G)f_{i1,u}+\sum_{k=0}^{m-1}u_{k+1}f_{i2,u_k}\right)dx\wedge dt+(f_{i1,u}+f_{i1,u_2})\,du_{2}\wedge dx
\\&+\sum_{\substack{k=1\\k\neq 2}}^{m} f_{i1,u_k}\,du_k\wedge dx+f_{i2,u_m}\,du_{m}\wedge dt.
\end{split}
\end{equation}
Given that the 1-forms $\omega_i$ satisfy the structure equation (\ref{2.10}), we substitute from (\ref{4.3}) to (\ref{4.4}). By setting the coefficients of the independent 2-forms to be zero, we derive
\begin{equation}\label{4.5}
f_{i1,u}+f_{i1,u_2}=0,
\end{equation}
\begin{equation}\label{4.6}
f_{i1,u_1}=f_{i1,u_k}=f_{i2,u_m}=0, \quad 1\leq i \leq 3,\quad 3\leq k \leq m,
\end{equation}
and
\begin{equation}\label{4.7}
-(\lambda u^{2}u_{3}+G)f_{11,u}+\sum_{k=0}^{m-1}u_{k+1}f_{12,u_k}+f_{32}f_{21}-f_{31}f_{22}=0,
\end{equation}
\begin{equation}\label{4.8}
-(\lambda u^{2}u_{3}+G)f_{21,u}+\sum_{k=0}^{m-1}u_{k+1}f_{22,u_k}+f_{12}f_{31}-f_{11}f_{32}=0,
\end{equation}
\begin{equation}\label{4.9}
-(\lambda u^{2}u_{3}+G)f_{31,u}+\sum_{k=0}^{m-1}u_{k+1}f_{32,u_k}+\delta(f_{12}f_{21}-f_{11}f_{22})=0.
\end{equation}
Taking the partial derivative of (\ref{4.7})-(\ref{4.9}) with respect to $ u_m, \ldots, u_4$ successively, we find that $ f_{i2,u_k} = 0 $ for all $ 3 \leq k \leq m $ and $ 1 \leq i \leq 3 $. Next, by taking the partial derivative with respect to $ u_3 $ for each equation as before, it follows from (\ref{3.4}) that
\begin{equation}\label{4.10}
f_{i2,u_2}=\lambda u^{2}f_{i1,u}=-\lambda u^{2}f_{i1,u_2},\quad 1\leq i \leq 3.
\end{equation}
Integrating in $ u_2 $, for each $ i $, there exists differentiable functions $ \phi_{i}(u,u_1) $ such that $ f_{i2} = -\lambda u^2 f_{i1} + \phi_{i}(u, u_1) $, $ 1 \leq i \leq 3 $. Therefore, we prove conditions (\ref{3.3})-(\ref{3.5}). Drawing on the hypotheses (\ref{3.2})-(\ref{3.5}), the equation (\ref{4.7}) reduces to
\begin{equation}\label{4.11}
-Gf_{11,u}+(-2\lambda u f_{11}-\lambda u^2 f_{11,u}+\phi_{1,u})u_{1}+\phi_{1,u_1}u_{2}
+(\mu_{2}\phi_{3}-\mu_{3}\phi_{2})f_{11}+\eta_{2}\phi_{3}-\eta_{3}\phi_{2}=0,
\end{equation}
which agrees with (\ref{3.6}). We substitute the equation (\ref{4.11}) into (\ref{4.8}) and (\ref{4.9}), and then we get (\ref{3.7}) and (\ref{3.8}), respectively. Note that condition (\ref{3.9}) ensures the existence of a metric defined on an open subset of $\mathbb{R}^2$. The converse computation is straightforward, and this completes the proof of the lemma.
$\hfill\square$
\end{Proof}	

\begin{Remark}
Under the conditions of Lemma 3.1, it follows from (\ref{3.4}) that the function $f_{11}$ satisfies $f_{11,u} + f_{11,u_2} = 0$. Hence, $f_{11}$ is a differentiable function in the variable $u - u_2$ such that $f_{11,u} = -f_{12,u_2} \neq 0$.
\end{Remark}

Next, we will prove Theorems 3.2-3.5 in Section 3.1 one by one. The analysis will be divided into two cases: $Q=0$ and $Q\neq 0$.

\hspace*{\fill}\\
\raggedright
\noindent\textbf{Case  $Q = 0$: Theorems 3.2-3.4}
\justifying

\setlength{\parindent}{2em}
We first present the following lemma, which can be derived by a straightforward calculation.

\begin{Lemma}
If $ Q = L_2=0 $, the system (\ref{3.7}) and (\ref{3.8}) can be equivalently expressed as
\begin{equation}\label{4.12}
L_{3}= 0,
\end{equation}
\begin{equation}\label{4.13}
\gamma\phi_{1} + 2\lambda\eta_2u u_1 = 0,
\end{equation}
\begin{equation}\label{4.14}
(-\mu_3N + \delta\eta_2)\phi_{1} - 2\lambda\eta_3u u_1= 0.
\end{equation}
\end{Lemma}

\noindent\textbf{4.2 Proof of Theorem 3.2}
\begin{Proof}
In Theorem 3.2, the functions $ f_{ij} $ satisfy conditions (\ref{3.1})-(\ref{3.10}), and we consider the case where $ Q = L_2 =0, \gamma = 0 $. Our goal is to determine the functions $ \phi_{i}(u,u_1) $, for $ 1 \leq i \leq 3$, satisfying (\ref{3.6})-(\ref{3.10}). Given that $ Q =L_2= 0 $, it follows from Lemma 4.2 that (\ref{3.7}) and (\ref{3.8}) are equivalent to (\ref{4.12})-(\ref{4.14}). Additionally, condition (\ref{3.9}) implies that $ \eta_2\phi_{1} \neq 0 $.
	
Using $ \gamma = 0 $ in (\ref{4.13}), we obtain $ \lambda\eta_2 = 0 $, which implies $\lambda = 0 $, and thus (\ref{4.14}) reduces to
\begin{equation}\label{4.15}
\frac{\delta(1+\mu_{2}^{2})-\mu_{3}^{2}}{1+\mu_{2}^2} \eta_{2}\phi_{1} =0,
\end{equation}
therefore $ \delta = 1 $ and $ 1 + \mu_2^2 = \mu_3^2 $. Given $ \gamma = 0 $, this leads to $ \eta_3 = \mu_2\eta_2/ \mu_3 $. Since $ \lambda = 0 $, it follows from (\ref{3.5}) that $ f_{12} = \phi_{1} $. Then, substituting $ L_2= 0 $ and (\ref{4.12}) into (\ref{3.6}), we determine the equation (\ref{3.1}) and associated 1-forms as described in Theorem 3.2, where we have introduced the notation $ f_{11} = f$. The converse computation is straightforward.
$\hfill\square$
\end{Proof}	

\noindent\textbf{4.3 Proof of Theorem 3.3}
\begin{Proof}
Since $L_2=0$ and $\gamma \neq 0$, it follows from (\ref{3.9}) and (\ref{4.13}) that $\eta_{2}\phi_{2}\neq 0$ and $\lambda \neq 0$ respectively. Moreover, (\ref{4.13}) implies that
\begin{equation}\label{4.16}
\phi_{1}=-\frac{2}{\gamma}\lambda \eta_{2}u u_{1},\quad \phi_{2}=-\frac{2}{\gamma}\lambda \mu_{2} \eta_{2}u u_{1},\quad \phi_{3}=-\frac{2}{\gamma}\lambda \mu_{3} \eta_{2}u u_{1}.
\end{equation}
Substituting $\phi_{1}$ into (\ref{4.14}), we get the identity $\delta\eta_{2}^{2}-\eta_{3}^{2}-(\mu_{2}\eta_{3}-\mu_{3}\eta_{2})^{2}=0$. We claim that $\delta=1$. In fact, if $\delta=-1$, then the identity imply $\eta_{2}=\eta_{3}=0$, which is contradictory to $\gamma\neq0$.

From (\ref{3.5}), we have $f_{i2}=-\lambda u^{2}f_{i1}+\phi_{i}(u,u_{1})$, where $\phi_{i}(u,u_1)$ is given by (\ref{4.16}).  And substituting (\ref{4.16}) into (\ref{3.6}), we conclude that the equation (\ref{3.1}) and associated 1-forms are given as in Theorem 3.3, where we also introduced the notation $f_{11}=f$. The converse is a straightforward computation.
$\hfill\square$
\end{Proof}

\begin{Lemma}
If $ Q = 0$ and $ L_2 \neq 0$, the system (\ref{3.7}) and (\ref{3.8}) can be equivalently expressed as
\begin{equation}\label{4.17}
L_3-\frac{\mu_{2}\mu_{3}}{1+\mu_{2}^{2}}L_2=0,
\end{equation}
\begin{equation}\label{4.18}
u_{1}L_{2,u}+u_{2}L_{2,u_1}-\gamma \phi_{1}-2\lambda \eta_{2}u u_{1}-\frac{\gamma \mu_{2}}{1+\mu_{2}^{2}}L_2=0,
\end{equation}
\begin{equation}\label{4.19}
\frac{\mu_{3}^{2}-\delta(1+\mu_{2}^{2})}{1+\mu_{2}^{2}}L_2f_{11}+\frac{\mu_{2}\mu_{3}}{1+\mu_{2}^{2}}(u_{1}L_{2,u}+u_{2}L_{2,u_1})
-\frac{\gamma \mu_{3}}{1+\mu_{2}^{2}}L_2-2\lambda \eta_{3}u u_{1}+(\mu_{2}\mu_{3}\eta_{3}-\mu_{3}^{2}\eta_{2}+\delta\eta_{2})\phi_{1}=0.
\end{equation}
\end{Lemma}

\noindent\textbf{4.4 Proof of Theorem 3.4}
\begin{Proof}
 In Theorems 3.4, the functions $f_{ij}$ satisfy (\ref{3.1})-(\ref{3.10}) and we are considering the case $Q=0$, $L_2\neq 0$ and $\gamma =0$. We want to determine the functions $\phi_{i}(u,u_1),1\leq i \leq 3 $ satisfying the system (\ref{3.6})-(\ref{3.10}). Since $Q =0$, $L_2\neq  0$, it follows from Lemma 4.3 that (\ref{3.7}) and (\ref{3.8}) are equivalent to (\ref{4.17})-(\ref{4.19}). Moreover, $\gamma =0$ by hypothesis, therefore (\ref{4.18}) reduced to
\begin{equation}\label{4.20}
u_{1}L_{2,u}+u_{2}L_{2,u_1}-2\lambda \eta_{2}u u_{1}=0.
\end{equation}
Hence,
\begin{equation}\label{4.21}
L_{2,u_1}=0,\quad L_2=\lambda \eta_{2}u^{2}+C_1,\quad (\lambda \eta_{2})^{2}+C_1^{2}=0,\quad C_1\in \mathbb{R},
\end{equation}
Substituting $L_2$ into (\ref{4.19}),we get
\begin{equation}\label{4.22}
\frac{\mu_{3}^{2}-\delta(1+\mu_{2}^{2})}{1+\mu_{2}^{2}}L_2f_{11}+(\mu_{2}\mu_{3}\eta_{3}-\mu_{3}^{2}\eta_{2}+\delta\eta_{2})\phi_{1}=0.
\end{equation}
Taking the derivative of above equation with respective to $u_2$, since $L_2f_{11,u_2}\neq 0$, we conclude that $\delta=1$ and $1+\mu_{2}^{2}=\mu_{3}^{2}$. Hence, the coefficient of $\phi_{1}$ in (\ref{4.22}) vanishes. From $\gamma =0$ we obtain $\eta_{3}=\pm\mu_{2}\eta_{2}/\sqrt{1+\mu_{2}^{2}}$. It follows from (\ref{3.6}) and (\ref{4.17}), that the equation (\ref{3.1}) and associated 1-forms are given as in Theorem 3,4, where we introduced the natation $f_{11}=f$. Since we have $(\lambda \eta_2)^{2}+C_1^{2} \neq 0$, condition (\ref{3.9}) is satisfied at once. The converse is a direct computation.

$\hfill\square$
\end{Proof}

In Theorem 3.4, we have only discussed the case $Q=0$, $ L_2\neq 0$ and $\gamma=0$. In fact, this is a matter of course because the case $\gamma \neq 0$ does not occur.
	
\begin{Lemma}
Assuming the conditions of Lemma 3.1 are satisfied, with $ Q = 0 $, $L_2 \neq 0 $, and $ \gamma \neq 0 $, then the system of equations (\ref{3.3})-(\ref{3.10}) has no solution.
\end{Lemma}
	
\begin{Proof}
Given the conditions of Lemma 3.1, the assertions of Lemma 4.3 are valid. Differentiating (\ref{4.18}) with respect to $ u_2 $ yields $ L_{2,u_1 }= 0 $ and
\begin{equation}\label{4.23}
\phi_{1}=\frac{1}{\gamma}\left(u_{1}L_{2,u}-2\lambda \eta_{2}u u_{1}-\frac{\gamma \mu_{2}}{1+\mu_{2}^{2}}L_2\right).
\end{equation}
Therefore, the equation (\ref{4.19}) rewrites to
\begin{equation}\label{4.24}
\begin{split}
\left(\frac{\mu_{2}\mu_{3}}{1+\mu_{2}^{2}}L_{2,u}-2\lambda \eta_{3}u-\frac{1}{\gamma}(\mu_{2}\mu_{3}\eta_{3}-\mu_{3}^{2}\eta_{2}+\delta\eta_{2})(L_{2,u}-2\lambda \eta_{2}u)\right)u_{1}
\\+\frac{\mu_{3}^{2}-\delta(1+\mu_{2}^{2})}{1+\mu_{2}^{2}}L_2f_{11}+\frac{\mu_{3}\eta_{3}-\delta\mu_{2}\eta_{2}}{1+\mu_{2}^{2}}L_2=0,
\end{split}
\end{equation}
Since $ L_2f_{11,u_2} \neq 0 $, differentiation of (\ref{4.24}) with respect to $ u_2 $ implies that $ \delta = 1 $ and $ \mu_2^3 = 1 + \mu_2^2 $. From $ L_{2,u_1} = 0 $, it follows that $ \mu_3\eta_3 - \mu_2\eta_2 = 0 $. Upon multiplying this result by $ \mu_3 $, we deduce that $ \gamma = 0 $, which contradicts the assumption that $ \gamma \neq 0 $.
$\hfill\square$
\end{Proof}

\hspace*{\fill}\\
\raggedright
\noindent\textbf{Case  $Q \neq 0$: Theorem 3.5}
\justifying

\setlength{\parindent}{2em}
Some lemmas will be put forward to prove this theorem.

\begin{Lemma}
Under the conditions of Lemma 3.1, if $Q\neq 0$, then (\ref{3.7}) and (\ref{3.8}) are equivalent to the following system
\begin{equation}\label{4.25}
f_{11}=\nu(u-u_2)-\sigma,
\end{equation}
\begin{equation}\label{4.26}
L_{2,u_1}=-\nu(L_{3}+\mu_{2}M),
\end{equation}
\begin{equation}\label{4.27}
u_{1}L_{2,u}-2\lambda \eta_{2}u u_{1}-\mu_{2}N+\eta_{3}\phi_{1}=\frac{(-\nu u+\sigma)} {\nu}L_{2,u_1},
\end{equation}
\begin{equation}\label{4.28}
\nu(\mu_{3}M+\delta L_2)+L_{3,u_1}=0,
\end{equation}
\begin{equation}\label{4.29}
(\nu u-\sigma)(\mu_{3}M+\delta L_2)-L_{3,u}u_{1}+2\lambda \eta_{3}u u_{1}+\mu_{3}N-\delta\eta_{2}\phi_{1}=0,
\end{equation}
where $\nu,\sigma\in \mathbb{R}$, with $\nu\neq 0$.
\end{Lemma}

\begin{Proof}
If $Q\neq 0$, then from (\ref{3.7}) we get the expression of $f_{11}$. Moreover, from (\ref{3.4}), we have that $f_{11,u}+f_{11,u_2}=0$. Hence
\begin{equation}\label{4.30}
\left(\frac{1}{Q}(u_{1}L_{2,u}-2\lambda \eta_{2}u u_{1}-\mu_{2}N+\eta_{3}\phi_{1})\right)_{u}+u_{2}\left(\frac{L_{2,u_1}}{Q}\right)_{u}+\frac{L_{2,u_1}}{Q}=0.
\end{equation}
Therefore,
\begin{equation}\label{4.31}
L_{2,u_1}=\nu Q,
\end{equation}
\begin{equation}\label{4.32}
\frac{1}{Q}(u_{1}L_{2,u}-2\lambda \eta_{2}u u_{1}-\mu_{2}N+\eta_{3}\phi_{1})=-\nu u+\sigma,
\end{equation}
where $\nu=\nu(u_1)$ and $\sigma=\sigma(u_1)$ are differentiable functions. Thus, (\ref{3.7}) implies that $f_{11}=\nu(u-u_2)-\sigma$. However, since $f_{11}$ does not depend on the variable $u_1$, it follows that $\nu,\sigma\in \mathbb{R}$ and $\nu\neq 0$. From (\ref{4.31}) and (\ref{4.32}), we obtain (\ref{4.26}) and (\ref{4.27}), respectively. Substituting $f_{11}$ into (\ref{3.8}), we obtain
\begin{equation}\label{4.33}
[\nu(\mu_{3}M+\delta L_2)+L_{3,u_1}]u_{2}-(\nu u-\sigma)(\mu_{3}M+\delta L_2)
+L_{3,u}u_{1}-2\lambda \eta_{3}u u_{1}-\mu_{3}N+\delta\eta_{2}\phi_{1}=0.
\end{equation}
Considering this equation as a polynomial in $u_2$, we get two equations that are equivalent to (\ref{4.28}) and (\ref{4.29}).
$\hfill\square$
\end{Proof}

\begin{Lemma}
Under the conditions of Lemma 3.1, let $Q \neq 0$ and $L_2=0 $, then the system of equations (\ref{3.3})-(\ref{3.10}) has no solution.
\end{Lemma}

\begin{Proof}
Since $Q \neq 0$ and $L_2=0$, we find $\phi_{2}=\mu_{2}\phi_{1}$ and $Q=-(1+\mu_{2}^{2})L_3$. Moreover, from (\ref{3.7}) we have
\begin{equation}\label{4.34}
-(1+\mu_{2}^{2})L_3f_{11}-2\lambda \eta_{2}u u_{1}-\mu_{2}N+\eta_{3}\phi_{1}=0.
\end{equation}
Taking the derivative of this equation with respect to $u_2$, since $f_{11,u_2}\neq 0$, we conclude that $L_3=0$ which leads to $Q=0$. This is a contradiction!
$\hfill\square$
\end{Proof}

Lemma 4.6 shows that, when $Q \neq 0$, the case $L_2=0 $ does not exist. Hence, it suffice to only consider the case $L_2\neq0 $. We introduce the following  useful notation for some constants
\begin{equation}\label{4.35}
\alpha := \delta(1+\mu_{2}^{2})-\mu_{3}^{2}, \quad \delta=\pm 1.
\end{equation}

\begin{Lemma}
Under the conditions of Lemma 3.1, if $Q\neq 0$ and $L_2\neq 0$, then the system of equations (\ref{3.6})-(\ref{3.8}) is equivalent to
\begin{equation}\label{4.36}
f_{11}=\nu(u-u_2)-\sigma,
\end{equation}
\begin{equation}\label{4.37}
G=\frac{1}{\nu}\left((-2\lambda u f_{11}+\phi_{1,u}-\lambda u^2 f_{11,u})u_{1}+\phi_{1,u_1}u_{2}+Mf_{11}+N\right),
\end{equation}
\begin{equation}\label{4.38}
\phi_{1}=\frac{1}{\gamma}\left(u_{1}L_{2,u}-2\lambda \eta_{2}u u_{1}-\frac{\mu_{2}\gamma}{1+\mu_{2}^{2}}L_2+\frac {1}{\nu}\left(\frac{\mu_{2}\eta_{2}}{1+\mu_{2}^{2}}+\nu u-\sigma \right)L_{2,u_1}\right),
\end{equation}
\begin{equation}\label{4.39}
\phi_{2}=L_2+\mu_{2}\phi_{1},
\end{equation}
\begin{equation}\label{4.40}
\phi_{3}=\mu_{3}\phi_{1}+\frac{\mu_{2}\mu_{3}}{1+\mu_{2}^{2}}L_2-\frac{L_{2,u_1}}{\nu(1+\mu_{2}^{2})},
\end{equation}
\begin{equation}\label{4.41}
L_{2,u_{1}u_{1}}-\nu^{2}\alpha L_2=0,
\end{equation}
\begin{equation}\label{4.42}
\begin{split}
[\mu_2\mu_3(\nu u-\sigma)+\mu_3\eta_2]\frac{L_{2,u_1}}{m}-[\alpha(\nu u-\sigma)+\mu_3\gamma]L_2+\left(\mu_2\mu_3 L_{2,u}-\frac{L_{2,u u_1}}{\nu}\right)u_1
\\ -2\lambda(1+\mu_2^2)\eta_3u u_1+(1+\mu_2^2)(\mu_2\mu_3\eta_3-\mu_3^2\eta_2+\delta\eta_2)\phi_{1}=0.
\end{split}
\end{equation}
\end{Lemma}

The proof is immediate from Lemma 4.5.
\hspace*{\fill}\\

To prove Theorem 3.5, the crucial step is to determine the functions $\phi_{i}(u,u_1)$, $1\leq i \leq 3$ satisfying (\ref{3.6})-(\ref{3.10}), which in this case by Lemma 4.7 are equivalent to (\ref{4.36})-(\ref{4.42}). We will divide two subcases, namely $\gamma=0$ and $\gamma \neq 0$ on an open set. Note that, in the first subcase, we can not obtain associated 1-forms satisfying the conditions, but the subcase $\gamma \neq 0$ does occur.

\begin{Lemma}
Under the conditions of Lemma 3.1, let $Q\neq 0$, $L_2\neq 0$ and $\gamma =0$, then the system of equations (\ref{3.3})-(\ref{3.10}) has no solution.
\end{Lemma}

\begin{Proof}
If $Q\neq 0$ and $L_2\neq 0$, it follows from Lemma 4,7 that (\ref{4.36})-(\ref{4.42}) holds. Next, we will prove that these equations have no solution. Since
$ \nu Q \neq 0$, then from (\ref{4.26}) we get
\begin{equation}\label{4.43}
L_{2,u_1}\neq 0.
\end{equation}
Moreover, since $\gamma =0$, (\ref{4.27}) implies
\begin{equation}\label{4.44}
u_{1}L_{2,u}-2\lambda \eta_{2}u u_{1}+\left(\frac{\mu_{2}\eta_{2}}{\nu(1+\mu_{2}^{2})}+\frac{\nu u-\sigma}{\nu}\right)L_{2,u_1}=0.
\end{equation}

\noindent There are now three subcases, depending on the sign of the constant $\alpha$:

If $\alpha <0$, then $L_2=c_{1}cos(\beta u_1)+c_{2}sin(\beta u_1)$, where $\beta :=|m|\sqrt{-\alpha}$ and $c_{1}=c_{1}(u)$ and $c_{2}=c_{2}(u)$ are differentiable functions. From (\ref{4.44}), we find
\begin{equation}\label{4.45}
\begin{split}
(c_{1}'cos(\beta u_1)+c_{2}'sin(\beta u_1))u_{1}-2\lambda \eta_{2}u u_{1}+\beta (c_{2}cos(\beta u_1)-c_{1}sin(\beta u_1))u
\\+\beta (c_{2}cos(\beta u_1)-c_{1}sin(\beta u_1))\left (\frac{\mu_{2}\eta_{2}}{\nu(1+\mu_{2}^{2})}-\frac{\sigma}{\nu}\right)=0,
\end{split}
\end{equation}
leading to $c_{1}=c_{2}=0$ and hence $L_2=0$, which contradicts the hypothesis $L_2\neq 0$. Thus, if $Q\neq 0$, $L_2\neq 0$,$\gamma =0$, and $\alpha <0$, then there are no solutions to the system of equations (\ref{4.36})-(\ref{4.42}).

If $\alpha=0$, then $L_2=c_{1}u_{1}+c_{2}$, where $c_{1}=c_{1}(u)$ and $c_{2}=c_{2}(u)$ are differentiable functions. Therefore, (\ref{4.44}) becomes to
\begin{equation}\label{4.46}
c_{1}'u_{1}^{2}+c_{2}'u_{1}-2\lambda \eta_{2}u u_{1}+c_{1}\left(\frac{\mu_{2}\eta_{2}}{\nu(1+\mu_{2}^{2})}+\frac{\nu u-\sigma}{\nu}\right)=0.
\end{equation}
Above equation implies that $c_{1}=c_{1}'=0$ and $c_{2}'=2\lambda \eta_{2}u$, and hence $L_{2,u_1}=0$ which contradicts (\ref{4.43}). Therefore, if $Q\neq 0$, $L_2\neq 0$,$\gamma =0$, and $\alpha =0$ then there are no solutions to the system of equations (\ref{4.36})-(\ref{4.42}).

If $\alpha >0$, then $L_2=c_{1}e^{\tau u_1}+c_{2}e^{-\tau u_1}$, where $\tau :=|m|\sqrt{\alpha}$ and $c_{1}=c_{1}(u)$ and $c_{2}=c_{2}(u)$ are differentiable functions. Therefore, we can rewrite (\ref{4.44}) as
\begin{equation}\label{4.47}
(c_{1}'e^{\tau u_1}+c_{2}'e^{-\tau u_1})u_{1}-2\lambda \eta_{2}u u_{1}+\tau (c_{1}e^{\tau u_1}-c_{2}e^{-\tau u_1})u
 +\tau (c_{1}e^{\tau u_1}-c_{2}e^{-\tau u_1})\left(\frac{\mu_{2}\eta_{2}}{\nu(1+\mu_{2}^{2})}-\frac{\sigma}{\nu}\right)=0.
\end{equation}
 Above equation also implies that $c_{1}=c_{2}=0$ and hence $L_2=0$, which contradicts the hypothesis $L_2\neq 0$. This mean that if $Q\neq 0$, $L_2\neq 0$,$\gamma =0$, and $\alpha >0$, then there are no solutions to the system of equations (\ref{4.36})-(\ref{4.42}).

In summary, whenever $Q\neq 0$, $L_2\neq 0$ and $\gamma =0$ the system of equations (\ref{3.3})-(\ref{3.10}) does not have any solution.
$\hfill\square$
\end{Proof}

\noindent\textbf{4.5 Proof of Theorem 3.5}
\begin{Proof}
Under the hypotheses of Theorem 3.5, Lemma 4.7 shows that the differential equation $u_{ t} - u_{2, t} = \lambda u^{2}u_3 + G $ can be characterized by the system of equations (\ref{4.36})-(\ref{4.42}). So by the same reasoning as above lemma, there exist three subcases based on the sign of the constant $\alpha$:

If $ \alpha < 0 $, then the system of equations (\ref{4.36})-(\ref{4.42}) has no solution. Specifically, for $ \alpha < 0 $, the solution of (\ref{4.41}) can be expressed as
\begin{equation}\label{4.48}
  L_2=c_{1}cos(\beta u_{1})+c_{2}sin(\beta u_{1}),
\end{equation}
where $ \beta :=|m| \sqrt{-\alpha}$, and $c_{1}=c_{1}(u)$ and $c_{2}=c_{2}(u)$  are differentiable functions of $ u $. As was done previously, an easy
induction shows that $ c_1 = c_2 = 0 $, and hence $ L_2 = 0 $. This leads to a contradiction.

If $ \alpha = 0 $, then $ \delta = 1 $, $1 + \mu_{2}^{2} = \mu_{3}^{2} $ and the solution of (\ref{4.41}) is provided by
\begin{equation}\label{4.49}
	L_2 = c_{1}u_{1} + c_{2},
\end{equation}
where $ c_{1} = c_{1}(u) $ and $c_{2} = c_{2}(u)$ are differentiable functions. From (\ref{4.38}) one has $ \phi_{1}(u,u_1) $ given in terms of $ c_{1} $, $ c_{2} $ and its derivatives and therefore (\ref{4.42}) reduces to
\begin{equation}\label{4.50}
  \left(\frac{c_{1}'}{\nu}-2\lambda \gamma u+\frac{\gamma c_1}{\mu_3}\right)u_{1}+\frac{1}{\mu_3}\left(\gamma c_{2}-\frac{\eta_{2}c_{1}}{\nu}\right)=0.
\end{equation}
By introducing the notation $ \theta := -\frac{\nu \gamma}{\mu_{3}} $, we find
\begin{equation}\label{4.51}
  c_{1}=\mu_{3}\left(\frac{2\lambda}{\theta}-\theta C_2e^{\theta u}+2\lambda u\right), \quad c_{2}=-\frac{\eta_{2}c_{1}}{\theta \mu_3},
\end{equation}
where $ C_2 \in \mathbb{R} $ and $\lambda^{2} + C_2^{2} \neq 0 $, since $ L_2 \neq 0 $. Thus,
\begin{equation}\label{4.52}
  L_2=\left(\frac{2\lambda}{\theta}-\theta C_2e^{\theta u}+2\lambda u\right)\left(\mu_{3}u_{1}-\frac{\eta_{2}}{\theta}\right),
\end{equation}
and (\ref{4.38}) reduces to
\begin{equation}\label{4.53}
  \phi_{1}=\frac{\mu_3}{\gamma}(2\lambda-\theta^{2}C_2e^{\theta u})u_{1}^{2}-\left(\frac{2\lambda}{\theta}-\theta C_2e^{\theta u}+2\lambda u\right)
\left(\left(\frac{\eta_2}{\gamma}+\frac{\mu_2}{\mu_3}\right)u_{1}+\frac{1}{\theta}(\nu u-\sigma)\right).
\end{equation}
Combining (\ref{4.39}) and (\ref{4.40}), we obtain
\begin{equation}\label{4.54}
  \phi_{2}=\mu_{2}\phi_{1}+\left(\frac{2\lambda}{\theta}-\theta C_2e^{\theta u}+2\lambda u\right)
\left(\mu_{3}u_{1}-\frac{\eta_{2}}{\theta}\right),
\end{equation}
\begin{equation}\label{4.55}
  \phi_{3}=\mu_{3}\phi_{1}+\left(\frac{2\lambda}{\theta}-\theta C_2e^{\theta u}+2\lambda u\right)
\left(\mu_{2}u_{1}-\frac{\eta_{3}}{\theta}\right).
\end{equation}
The results $1 + \mu_2^2 = \mu_3^2 $ and $ \theta\mu_3 + \nu\gamma = 0 $ yield
\begin{equation}\label{4.56}
  \eta_{3}=\pm\frac{\theta+\mu_{2}\eta_{2}\nu}{\nu\sqrt{1+\mu_{2}^{2}}}.
\end{equation}
Let
\begin{equation}\label{4.57}
  \zeta_{1}=\frac{2\sigma}{\nu}-\frac{1}{\theta}-\frac{\theta}{\nu^{2}(1+\mu_{2}^{2})}-\frac{\eta_{2}(2\theta \mu_{2}+\nu\eta_{2})}{\theta \nu(1+\mu_{2}^{2})}.
\end{equation}
Using (\ref{3.1}) and referring to (\ref{3.5}) and (\ref{4.37}), we deduce that the equation (\ref{3.1}) and associated 1-forms as demonstrated in Theorem 3.5(i). The converse is derived through direct computation.
	
If $ \alpha > 0 $, then $ \delta = 1 $, $1 + \mu_{2}^{2} = \mu_{3}^{2} $ and the solution of (\ref{4.41}) is
\begin{equation}\label{4.58}
  L_2=c_{1}e^{\tau u_{1}}+c_{2}e^{-\tau u_{1}},
\end{equation}
where $ \tau :=|m|\sqrt{\alpha}$, and $ c_1 = c_1(u) $ and $ c_2 = c_2(u) $ are differentiable functions. Consequently, utilizing (\ref{4.38}), we obtain
\begin{equation}\label{4.59}
  \begin{split}
\phi_{1}&=\frac{1}{\gamma}\left(c_{1}'u_{1}-\frac{\mu_{2}\gamma}{1+\mu_{2}^{2}}c_{1}
+\left(\frac{\mu_{2}\eta_{2}}{\nu(1+\mu_{2}^{2})}+\frac{\nu u-\sigma}{\nu}\right)\tau c_{1}\right)e^{\tau u_1}
\\&+\frac{1}{\gamma}\left(c_{2}'u_{1}-\frac{\mu_{2}\gamma}{1+\mu_{2}^{2}}c_{1}
-\left(\frac{\mu_{2}\eta_{2}}{\nu(1+\mu_{2}^{2})}+\frac{\nu u-\sigma}{\nu}\right)\tau c_{2}\right)e^{-\tau u_1}-\frac{2\lambda \eta_{2}}{\gamma}u u_{1}.
\end{split}
\end{equation}
Substituting $\phi_{1}$ and $L_2$ into (\ref{4.42}), we get
\begin{equation}\label{4.60}
(c_{1}'u_{1}+c_{1}P(u))\xi^{-}e^{\tau u_1}+(c_{2}'u_{1}-c_{2}P(u))\xi^{+}e^{-\tau u_1}+2\lambda \left(\gamma -\frac{\alpha \eta_{2}^{2}}{\gamma}\right)u u_{1}=0,
\end{equation}
where
\begin{equation}\label{4.61}
  P(u)=\left(u-\frac{\sigma}{\nu}\right)\tau +\frac{\nu(\mu_{3}\gamma +\alpha \mu_{2}\eta_{2})}{\tau(1+\mu_{2}^{2})},
\end{equation}
and $\xi^{\pm}$ are constants defined by
\begin{equation}\label{4.62}
  \xi^{\pm}=\frac{\alpha \eta_{2}}{\gamma}\pm \frac{\tau}{\nu}.
\end{equation}
By (\ref{4.60}) and the condition $P(u)\neq 0$, the following identities must hold
\begin{equation}\label{4.63}
c_{1}'\xi^{-}=0,\quad c_{2}'\xi^{+}=0,\quad c_{1}\xi^{-}=0,\quad c_{2}\xi^{+}=0,\quad \lambda \left(\gamma -\frac{\alpha \eta_{2}^{2}}{\gamma}\right)=0.
\end{equation}
Note that $\xi^{\pm}=0$ implies $\gamma=\mp\frac{\alpha \nu\eta_{2}}{\tau}$, which means $\gamma^{2}=\alpha \eta_{2}^{2}$. Furthermore, since $\alpha >0$ we see
\begin{equation}\label{4.64}
  (\xi^{+})^{2}+(\xi^{-})^{2}=2\alpha\left(1+\frac{\alpha \eta_{2}^{2}}{\gamma}\right)\neq 0.
\end{equation}
Therefore, from (\ref{4.63}) and $ L_2 \neq 0 $, we deduce that either $c_{1}= \xi^{+}=0$ or $c_{2}=\xi^{-}=0$. In both cases, $\gamma \neq 0$ implies that $\eta_{2}\neq 0$. Consequently, we consider $L_2$ as
\begin{equation}\label{4.65}
  L_2=\pm \eta_{2}\tau\varphi(u)e^{\pm\tau u_1},
\end{equation}
where $\varphi(u)$ is a differentiable function of $u$ and $\varphi \neq 0$. Using (\ref{4.39}), (\ref{4.40}) and (\ref{4.59}), we have
\begin{equation}\label{4.66}
\begin{split}
&\phi_{1}=[\pm\tau(\nu u-\sigma)\varphi +\nu\varphi'u_{1}]e^{\pm\tau u_1}\mp\frac{2\lambda \nu}{\tau}u u_{1},
\\&\phi_{2}=\mu_{2}\phi_{1}\pm \eta_{2}\tau\varphi e^{\pm\tau u_1},
\\&\phi_{3}=\mu_{3}\phi_{1}\pm \eta_{3}\tau\varphi e^{\pm\tau u_1}.
\end{split}
\end{equation}
Let
\begin{equation}\label{4.67}
		 \zeta_{2}=\frac{\sigma}{\nu}\mp\frac{\mu_{3}\eta_{2}-\mu_{2}\eta_{3}}{\tau}.
\end{equation}
Since $ \gamma^2 = \alpha\eta_2^2 $, it follows from (\ref{4.35}) that $ \mu_3 $ and $ \eta_3 $ can be rewritten as
\begin{equation}\label{4.68}
  \mu_{3}=\pm\tau \left(\frac{1+\mu_{2}^{2}}{\eta_2}(\frac{\sigma}{\nu}-\zeta_{2})-\frac{\mu_{2}}{\nu}\right),
\end{equation}
\begin{equation}\label{4.69}
  \eta_{3}=\pm\tau \left(\mu_{2}(\frac{\sigma}{\nu}-\zeta_{2})-\frac{\eta_{2}}{\nu}\right).
\end{equation}
We conclude from (\ref{3.5}), (\ref{4.37}) and the above expressions that the equation (\ref{3.1}) and associated 1-forms are given as in Theorem 3.5(ii). The converse is a straightforward computation. This completes the proof of Theorem 3.5.
$\hfill\square$
\end{Proof}
\section{Concluding remarks}
\indent \indent In this paper, we consider a class of third order differential equations with the form
\begin{equation}\label{5.1}
u_t-u_{xxt}=\lambda u^2u_{xxx}+G(u, u_x, u_{xx}).
\end{equation}
Under certain assumption on the coefficients of the connection 1-form associated to the surfaces, we are able to give a classification of \eqref{5.1} that describes a pseudoshperical or spherical surface. As examples, we show that series of equations belong to such class, such as gCH equation, and equations such as the Novikov equation does not describe pseudoshperical surfaces.

Systems of CH-type have natural geometric correspondence, such as the multi-component CH system \cite{klq21}. It is worthwhile to see whether the argument used in this paper can be applied to a more general cases than \eqref{5.1}. Therefore, we can find  more CH-type systems which can be connected to pseudospherical or spherical surfaces. Meanwhile, the sine-Gordon equation and the B\"{a}cklund theorem can be generalized to higher dimension \cite{TT80, T80}. One natural question to ask is how to construct Darboux-B\"acklund transformation for these CH type equations, which will lead to geometric transformation of associated pseudospherical surfaces.
\section*{Acknowledgements} Guo's research is supported by Northwest University Graduate Research and Innovation Program CX2024133. Kang's research is supported by NSFC (Grant No. 12371252) and Basic Science Program of Shaanxi Province (Grant No. 2019JC-28). Shi's research is supported by NSFC (Grant No.11901454). Wu's research is supported by NSFC (Grant No. 12271535 and 12431008)


\begin{thebibliography}{99}
\bibitem{ablowitz1974}
M. J. Ablowitz, D. J. Kaup, A. C. Newell, and H. Segur, \emph{The inverse scattering transform-Fourier analysis for nonlinear problems}, Studies in Applied Mathematics \textbf{53} (1974): 249-315.

\bibitem{bour1891}E. Bour, \emph{Th\'eorie de la d\'eformation des surfaces}, Journal de l'\'Ecole polytechnique. Math\'ematiques  \textbf{19} (1862): 1-48.

\bibitem{silva2015}
T. Castro Silva and K. Tenenblat, \emph{Third order differential equations describing pseudospherical surfaces}, Journal of Differential Equations \textbf{259} (2015): 4897-4923.

\bibitem{ferraioli2024}
D. Catalano Ferraioli and T. Castro Silva, \emph{A class of third order quasilinear partial differential equations describing spherical or pseudospherical surfaces},  Journal of Differential Equations \textbf{379} (2024): 524-568.

\bibitem{ferraioli2020}
D. Catalano Ferraioli, T.  Castro Silva, and K. Tenenblat,  \emph{A class of quasilinear second order partial differential equations which describe spherical or pseudospherical surfaces}, Journal of Differential Equations \textbf{268} (2020): 7164-7182.

\bibitem{ferraioli2016}
D. Catalano Ferraioli and L. A. de Oliveira Silva, \emph{Second order evolution equations which describe pseudospherical surfaces}, Journal of Differential Equations \textbf{260} (2016): 8072-8108.

\bibitem{ferraioli2014}
D. Catalano Ferraioli and K. Tenenblat, \emph{Fourth order evolution equations which describe pseudospherical surfaces}, Journal of Differential Equations \textbf{257} (2014): 3165-3199.

\bibitem{cavalcante1988}
 J. Cavalcante and K. Tenenblat, \emph{Conservation laws for nonlinear evolution equations}, Journal of Mathematical Physics \textbf{29} (1988): 1044-1049.

\bibitem{chern1986}S. S. Chern and K. Tenenblat, \emph{Pseudospherical surfaces and evolution equations},  Studies in Applied Mathematics \textbf{74} (1986): 55-83.

\bibitem{crampin1977}
 M. Crampin, F. Pirani, and D. Robinson, \emph{The soliton connection}, Letters in Mathematical Physics \textbf{2} (1977): 15-19.

\bibitem{ding2002}
Q. Ding and K. Tenenblat, \emph{On differential systems describing surfaces of constant curvature}, Journal of Differential Equations \textbf{184} (2002): 185-214.

\bibitem{neto2010}
 V. P. Gomes Neto, \emph{Fifth order evolution equations describing pseudospherical surfaces}, Journal of Differential Equations \textbf{249} (2010): 2822-2865.

\bibitem{jorge1987}
 L. Jorge and K. Tenenblat, \emph{Linear problems associated with evolution equations of the type $u_{tt} = F(u, u_x, u_{xx}, u_t)$}, Studies in Applied Mathematics \textbf{77} (1987): 103-117.

\bibitem{kamran1995}
N. Kamran and K. Tenenblat, \emph{On differential equations describing pseudospherical surfaces}, Journal of Differential Equations \textbf{115} (1995): 75-98.

\bibitem{klq21}
J. Kang, X. Liu and C. Qu, \emph{On an integrable multi-component Camassa-Holm system arising from M\"obius geometry}, Proceedings A \textbf{477} (2021): 20210164.

\bibitem{kelmer2022}
F. Kelmer and K. Tenenblat, \emph{On a class of systems of hyperbolic equations describing pseudospherical or spherical surfaces}, Journal of Differential Equations \textbf{339} (2022): 372-394.

\bibitem{kelmer2025}
F. Kelmer and K. Tenenblat, \emph{Systems of differential equations of higher order describing pseudo-spherical or spherical surfaces}, Journal of Differential Equations \textbf{424} (2025): 833-858.

\bibitem{novikov2009}
V. Novikov, \emph{Generalizations of the Camassa-Holm equation}, Journal of Physics A-Mathematical and Theoretical \textbf{42} (2009): 342002.


\bibitem{rabelo1989}
M. L. Rabelo, \emph{On equations which describe pseudospherical surfaces},  Studies in Applied Mathematics \textbf{81} (1989): 221-248.


\bibitem{rabelo1992}
 M. L. Rabelo and K. Tenenblat,  \emph{A classification of pseudospherical surface equations of type $u_t = u_ {xxx} + G(u, u_x, u_{xx})$}, Journal of Mathematical Physics  \textbf{33} 1992: 537-549.

 \bibitem{reyes1998}
E. G. Reyes, \emph{Pseudospherical surfaces and integrability of evolution equations}, Journal of Differential Equations \textbf{147} (1998): 195-230.

\bibitem{reyes2002}
E. G. Reyes, \emph{Geometric integrability of the Camassa-Holm equation}, Letters in Mathematical Physics \textbf{59} (2002): 117-131.

\bibitem{reyes2003}
E. G. Reyes, \emph{On generalized B\"acklund transformations for equations describing pseudospherical surfaces}, Journal of Geometry and Physics \textbf{45} (2003): 368-392.

\bibitem{reyes2005}
E. G. Reyes, \emph{Nonlocal symmetries and the Kaup-Kupershmidt equation}, Journal of Mathematical Physics \textbf{46} (2005): 073507.

\bibitem{reyes2005b}
E. G. Reyes, \emph{The soliton content of classical Jackiw-Teitelboim gravity},  Journal of Physics A: Mathematical and General \textbf{39} (2005): 55-63.

\bibitem{reyes2006}
E. G. Reyes, \emph{Pseudo-potentials, nonlocal symmetries and integrability of some shallow water equations}, Selecta Mathematica \textbf{12} (2006): 241-270.

\bibitem{reyes2011}
E. G. Reyes, \emph{Equations of pseudo-spherical type (After S. S. Chern and K. Tenenblat)}, Results in Mathematics \textbf{60} (2011): 53-101.

\bibitem{sasaki1979}R. Sasaki, \emph{Soliton equations and pseudospherical surfaces}, Nuclear Physics B \textbf{154} (1979): 343-357.

\bibitem{TT80}
K. Tenenblat and C.L. Terng, \emph{B\"{a}cklund's theorem for n-dimensional submanifolds of $\mathbb{R}^{2n-1}$}, Annals of Mathematics \textbf{111} (1980): 477-490.

\bibitem{T80}
C.L. Terng, \emph{A higher dimension generalization of the Sine-Gordon equation and its soliton theory}, Annals of Mathematics \textbf{111} (1980): 419-510.

\bibitem{wadati1979}
 M. Wadati, K. Konno, and Y. H. Ichikawa, \emph{A generalization of inverse scattering method}, Journal of the Physical Society of Japan \textbf{6} (1979): 1965-1966.






\end{thebibliography}
\end{document}